\documentclass{article} 
\usepackage{iclr2024_conference_tinypaper,times}

\usepackage{amsmath,amsfonts,bm}

\def\eqref#1{equation~\ref{#1}}

\def\1{\bm{1}}

\DeclareMathAlphabet{\mathsfit}{\encodingdefault}{\sfdefault}{m}{sl}
\SetMathAlphabet{\mathsfit}{bold}{\encodingdefault}{\sfdefault}{bx}{n}

\usepackage{hyperref}
\usepackage{url}
\usepackage{graphicx}
\usepackage{float}
\usepackage{multirow}
\usepackage{comment}
\usepackage{subcaption}

\title{A Novel Sector-Based Algorithm for an Optimized Star-Galaxy Classification}

\author{Anuamnchi Agastya Sai Ram Likhit, Divyansh Tripathi, Akshay Agarwal \\
Department of Physics, Department of Data Science and Engineering\\
Indian Institute of Science Education and Research, Bhopal\\
\texttt{\{anumanchi20, divyansh20, akagarwal\}@iiserb.ac.in}
}

\iclrfinalcopy 
\begin{document}

\maketitle

\begin{abstract}
This paper introduces a novel sector-based methodology for star-galaxy classification, leveraging the latest Sloan Digital Sky Survey data (SDSS-DR18). By strategically segmenting the sky into sectors aligned with SDSS observational patterns and employing a dedicated convolutional neural network (CNN), we achieve state-of-the-art performance for star galaxy classification. Our preliminary results demonstrate a promising pathway for efficient and precise astronomical analysis, especially in real-time observational settings.

\end{abstract}

\section{INTRODUCTION}
\begin{figure}[b]
\centering
\includegraphics[width = 0.85\linewidth]{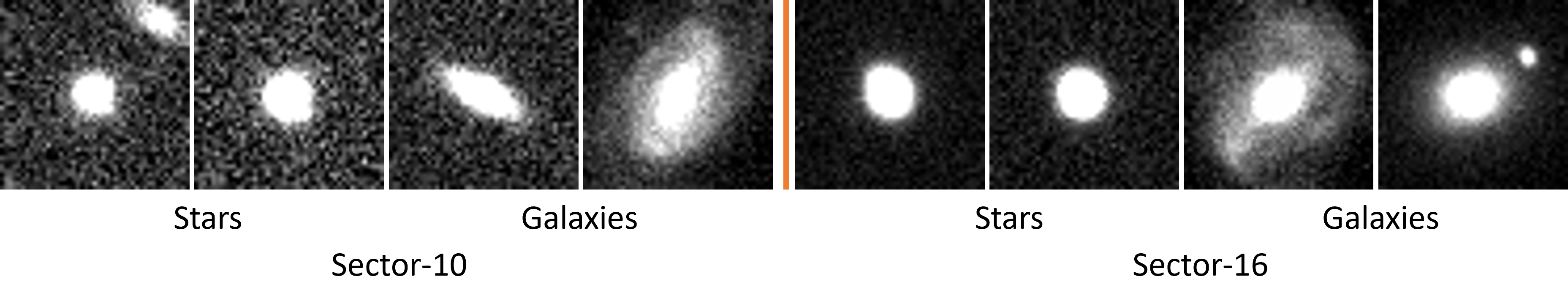}
\vspace{-5mm}
\caption{A sample image reflecting the challenges in identifying star-galaxies in different sectors.}
\label{samples}
\end{figure}

\vspace{-3mm}
Today is the age of Data-Driven astronomy, with sky surveys generating large amounts of data, and many new ones are lining up, such as the large synoptic survey telescope (LSST). One of the key motives of such surveys is to classify objects as stars or galaxies. However, manual classification can not be done for petabytes of data and large intra-class variation, which raises the need for an automated and robust classification model. Recently, several research works have been developed to help astronomers by automatically classifying the galaxies \citep{10.1093/mnras/stu1410,Alawiinproceedings,Chaini_2022,Kim_2016,9952065}. However, these models perform well but are complex. In contrast to the existing work, due to the complexity of our star-galaxy system, in this research, we have proposed the development of a classification approach utilizing a sector-based division of the sky. The prime motivation for such division can be seen in Figure \ref{samples}, reflecting the variation present in different sectors and difficulties in classification. By utilizing these differences, we have developed a star-galaxy classification system that surpasses existing algorithms and yields a low computational cost.

\section{Proposed Methodology}\label{gen_inst}
\vspace{-2mm}
To address the star-galaxy classification challenge, we introduce a sector-based approach closely aligned with the Sloan Digital Sky Survey (SDSS)\citet{SDSSDR18_2023} observation patterns. For that sky is divided into thirty-six distinct sectors (Appendix \ref{appen-1} \(\&\) Figure \ref{fig:sectors}) by segmenting Right Ascension (RA) and Declination (Dec) intervals. \textit{Right Ascension (RA) is akin to longitudinal lines on Earth and ranges from 0 hours to $24$ hours, equivalent to $0^o$ to $360^o$ in celestial coordinates}. RA is divided into six equal intervals of $60^o$ each, corresponding to $4$-hour segments. Declination (Dec) which spans from the North celestial pole at $+90^o$ to the South celestial pole at $-90^o$ is segmented into six intervals of $30^o$ each. Combining the divisions in RA and Dec, we obtain a total of 36 sectors which are defined by their specific RA and Dec range\footnote{The other necessary details are also provided in the Appendix.}. 

Once the divided sector images are obtained, we have provided them as input to the proposed custom convolutional neural network model  (Appendix \ref{appen-2}). As mentioned above, we asserted that the division of the sky into sectors can better help in star-galaxy classification even with the use of a simple model; therefore, a shallow model of $3$ convolutional layers have been developed. The aim is not only to achieve higher accuracy but at a lower computational cost. Each convolutional layer is followed by max-pooling and ReLU activation functions. Two dense layers each containing $64$ neurons are attached at the end to extract the features. The dropout is also added to reduce the impact of overfitting. The configuration of the proposed CNN is given in Table \ref{tab:CNN-Arch} (Appendix \ref{appen-2}).

\section{EXPERIMENTS AND RESULTS}
\vspace{-2mm}

\begin{table}[t]
\caption{Star-Galaxy Classification performance of the proposed and existing Algorithms in terms of accuracy (Acc), precision (P), recall (R), and F-1 score (F1).}
\label{tab:accuracy}
\vspace{-3mm}
\resizebox{\textwidth}{!}{%
\begin{tabular}{|c|cccc|cccc|cccc|}
\hline
\multirow{2}{*}{Algorithm} &
  \multicolumn{4}{c|}{Sector 10} &
  \multicolumn{4}{c|}{Sector 16} &
  \multicolumn{4}{c|}{Combined} \\ \cline{2-13}
&
  \multicolumn{1}{l|}{Acc} &
  \multicolumn{1}{l|}{P} &
  \multicolumn{1}{l|}{R} &
  \multicolumn{1}{l|}{F1} &
  \multicolumn{1}{l|}{Acc} &
  \multicolumn{1}{l|}{P} &
  \multicolumn{1}{l|}{R} &
  \multicolumn{1}{l|}{F1} &
  \multicolumn{1}{l|}{Acc} &
  \multicolumn{1}{l|}{P} &
  \multicolumn{1}{l|}{R} &
  \multicolumn{1}{l|}{F1} \\ \hline \hline
\textbf{Proposed} &
  \multicolumn{1}{c|}{\textbf{0.96}} &
  \multicolumn{1}{c|}{\textbf{0.97}} &
  \multicolumn{1}{c|}{\textbf{0.97}} &
  \textbf{0.97} &
  \multicolumn{1}{c|}{\textbf{0.95}} &
  \multicolumn{1}{c|}{\textbf{0.95}} &
  \multicolumn{1}{c|}{\textbf{0.95}} &
  \textbf{0.95} &
  \multicolumn{1}{c|}{\textbf{0.95}} &
  \multicolumn{1}{c|}{\textbf{0.95}} &
  \multicolumn{1}{c|}{\textbf{0.95}} &
  \textbf{0.95} \\ \hline\hline
CovNet &
  \multicolumn{1}{c|}{0.91} &
  \multicolumn{1}{c|}{0.91} &
  \multicolumn{1}{c|}{0.91} &
  0.91 &
  \multicolumn{1}{c|}{0.94} &
  \multicolumn{1}{c|}{0.94} &
  \multicolumn{1}{c|}{0.94} &
  0.94 &
  \multicolumn{1}{c|}{0.88} &
  \multicolumn{1}{c|}{0.89} &
  \multicolumn{1}{c|}{0.89} &
  0.89 \\ \hline\hline
MargNet &
  \multicolumn{1}{c|}{0.94} &
  \multicolumn{1}{c|}{0.94} &
  \multicolumn{1}{c|}{0.94} &
  0.94 &
  \multicolumn{1}{c|}{0.93} &
  \multicolumn{1}{c|}{0.94} &
  \multicolumn{1}{c|}{0.94} &
  0.94 &
  \multicolumn{1}{c|}{0.92} &
  \multicolumn{1}{c|}{0.93} &
  \multicolumn{1}{c|}{0.92} &
  0.92 \\ \hline
\end{tabular}%
}
\vspace{-3mm}
\end{table}

\begin{table}[t]
\centering
\caption{Confusion matrix reflecting the effectiveness in handling individual sectors and classes.}
\label{cm}
\vspace{-3mm}
\begin{tabular}{|l|l||l|c||l|c||l|c|}
\hline
\multirow{2}{*}{\begin{tabular}[c]{@{}l@{}}Algorithm \end{tabular}} & \multirow{2}{*}{\begin{tabular}[c]{@{}l@{}}True $\downarrow$\\ Predicted $\rightarrow$\end{tabular}} & \multicolumn{2}{l||}{Sector-10} & \multicolumn{2}{l||}{Sector-16} & \multicolumn{2}{l|}{Combined} \\ \cline{2-8} 
& & \multicolumn{1}{l|}{Star} & \multicolumn{1}{l||}{Galaxy} & \multicolumn{1}{l|}{Star} & \multicolumn{1}{l||}{Galaxy} & \multicolumn{1}{l|}{Star} & \multicolumn{1}{l|}{Galaxy} \\ \hline

\multirow{2}{*}{\begin{tabular}[c]{@{}l@{}}\textbf{\underline{Proposed}} \end{tabular}} & Star & \multicolumn{1}{c|}{\textbf{935}} & 53 & \multicolumn{1}{c|}{\textbf{963}} & 25 & \multicolumn{1}{c|}{\textbf{1858}} & 123 \\ \cline{2-8} 
& Galaxy & \multicolumn{1}{c|}{44} & \textbf{968} & \multicolumn{1}{c|}{44} & \textbf{968} & \multicolumn{1}{c|}{67} & \textbf{1952} \\ \hline

\multirow{2}{*}{\begin{tabular}[c]{@{}l@{}}CovNet \end{tabular}} & Star & \multicolumn{1}{c|}{\textbf{894}} & 89 & \multicolumn{1}{c|}{\textbf{918}} & 70 & \multicolumn{1}{c|}{\textbf{1742}} & 238 \\ \cline{2-8}
& Galaxy & \multicolumn{1}{c|}{85} & \textbf{927} & \multicolumn{1}{c|}{42} & \textbf{970} & \multicolumn{1}{c|}{217} & \textbf{1802} \\ \hline

\multirow{2}{*}{\begin{tabular}[c]{@{}l@{}}MargNet \end{tabular}} & Star & \multicolumn{1}{c|}{\textbf{952}} & 36 & \multicolumn{1}{c|}{\textbf{907}} & 81 & \multicolumn{1}{c|}{\textbf{1933}} & 48 \\ \cline{2-8}
& Galaxy & \multicolumn{1}{c|}{76} & \textbf{936} & \multicolumn{1}{c|}{44} & \textbf{968} & \multicolumn{1}{c|}{268} & \textbf{1751} \\ \hline

\end{tabular}
\vspace{-5mm}
\end{table}


The primary focus of the evaluation is on sector-$10$ and sector-$16$ \footnote{More details on the selection of these sectors are provided in the appendix \ref{appen-4}}. Each sector contributed $10,000$ images and the proposed classification model is trained separately on each sector to determine their effectiveness in handling individual sectors. Moreover, to evaluate the scalability of the proposed algorithm is also evaluated on the large-scale dataset achieved after combining the images of both sectors. Further, a comparison with existing SOTA algorithms: CovNet \citet{Kim_2016} and MargNet \citet{Chaini_2022} has also been performed to demonstrate the efficacy of the proposed approach. For a fair comparison, the existing models are trained on the same training-testing setting on which the proposed algorithm is trained.

The comparative results reported in Table \ref{tab:accuracy} demonstrate that the proposed algorithm surpasses each existing algorithm by a significant margin. Further, the effectiveness of the proposed algorithm is not only for a single sector but also for each and in combination. For example, the proposed algorithm on a large-scale dataset comprising sector-10 and sector-16 yields an accuracy of $\mathbf{95.25}$\% in comparison to $88.62$\% and $92.10$\% of CovNet and MargNet, respectively. Further, Table \ref{cm} shows the confusion matrix of the proposed algorithm. It shows that the proposed algorithm is not biased to any particular class and can effectively identify stars and galaxies with higher accuracy. 

\textit{As shown in Table \ref{tab:RunningTime} of the Appendix \ref{appen-2}, the proposed algorithm took $\mathbf{25}$s/epoch on the data of the combined sectors as compared to $180$sec and $1610$sec taken by CovNet and MargNet, respectively}. 

\vspace{-2mm}
\section{CONCLUSION}
\vspace{-4mm}
We have proposed a novel and cost-effective algorithm for star-galaxy classification by handling sector-specific data. The efficacy of the proposed algorithm surpasses the existing algorithm back our idea of segregating the sky into sectors for better performance. In the future, we aim to develop an advanced architecture to tackle other sectors and improve the classification performance of the proposed approach by incorporating sector-specific auxiliary information. We believe the proposed research can advance the astronomical research by precisely identifying the celestial objects.

\subsubsection*{URM Statement}
The authors acknowledge that the key author of this work meets the URM criteria of the ICLR 2024 Tiny Papers Track.

\bibliography{iclr2024_conference_tinypaper}
\bibliographystyle{iclr2024_conference_tinypaper.bst}

\appendix

\section{Sky Division}\label{appen-1}

We want to highlight that this division aligns with the inherent structure of the SDSS, which scans the sky in stripes spaced $2.5^o$ apart in survey latitude. Using a 30° sector in declination, each sector encompasses 12 SDSS stripes. This alignment ensures data homogeneity within each sector, containing a complete and consistent set of SDSS stripes.

The 60° x 30° sector division strikes a balance between granularity and expansive sky coverage. It provides a detailed sky view while maintaining enough breadth for comprehensive sector-specific analysis. This division ensures that the data for each sector is substantial yet not overly dense, allowing computational algorithms to be applied effectively on a per-sector basis without requiring excessive resources.
 
Moreover, this sector division is not only tailored for the SDSS data structure but also remains universally applicable. The equatorial coordinate-based division offers a flexible foundation to integrate additional survey data in the future. Additionally, the polar regions, often challenging in sky segmentation due to projection distortion, are effectively managed in this division. The sectors that extend from +60° to +90° and -60° to -90° handle the polar regions without significant distortion.

\begin{figure}[h]
    \centering
    \begin{subfigure}[b]{0.45\textwidth} 
        \centering
        \includegraphics[scale=0.23]{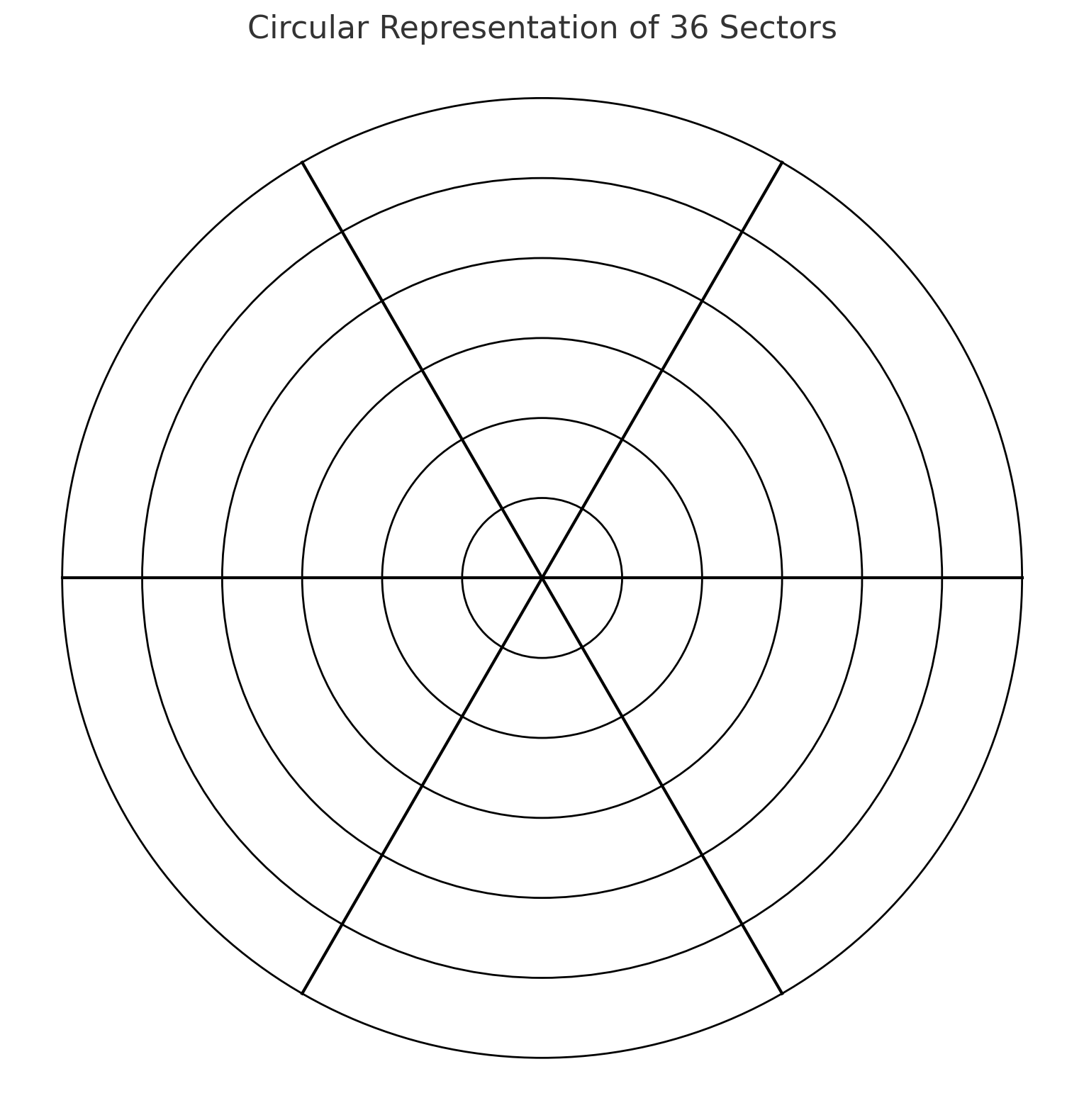} 
        \caption{Circular Representation}
        \label{fig:sub1}
    \end{subfigure}
    \hfill 
    \begin{subfigure}[b]{0.45\textwidth} 
        \centering
        \includegraphics[scale=0.23]{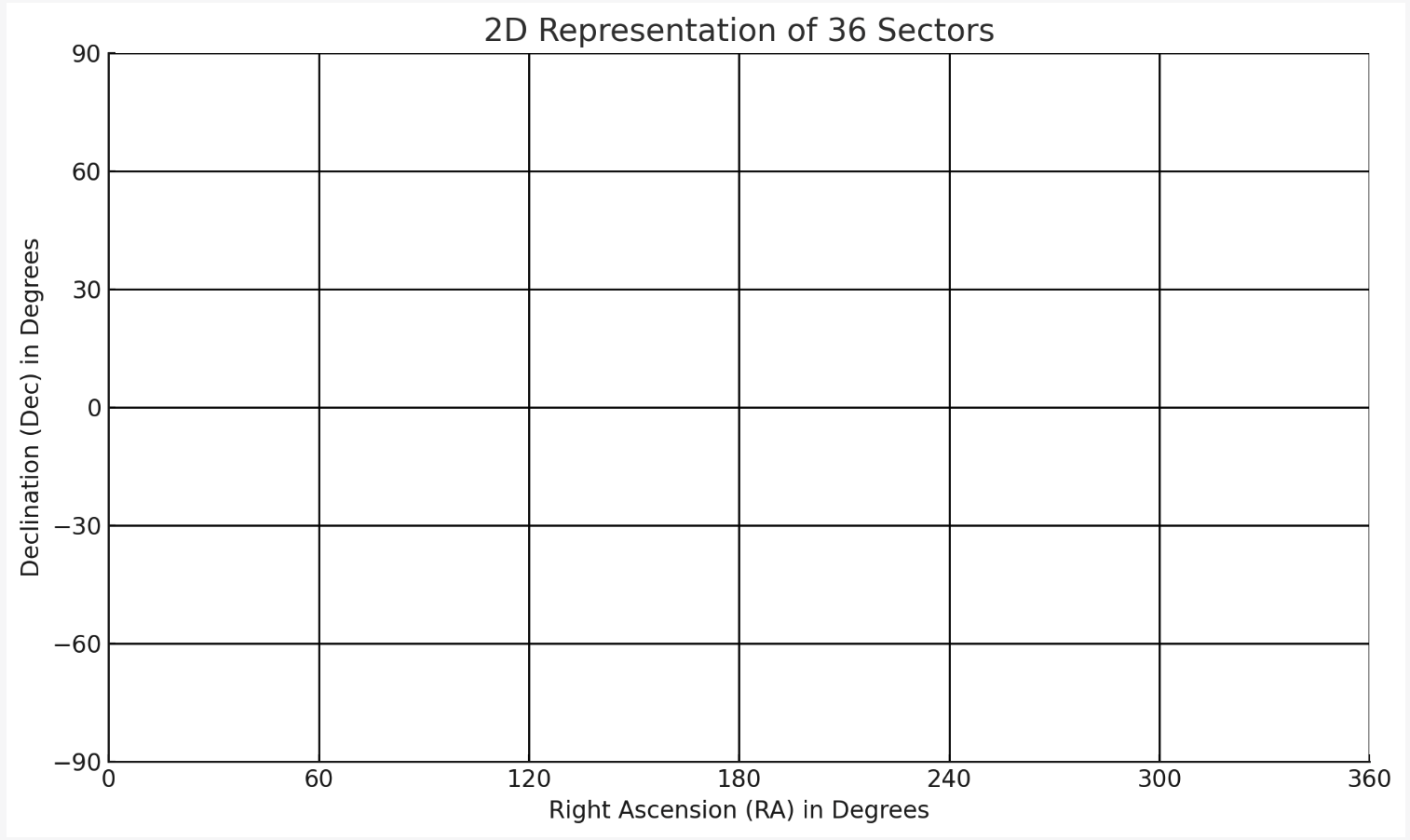} 
        \caption{Rectangular Representation}
        \label{fig:sub2}
    \end{subfigure}
    \caption{2D Sky Sector Map}
    \label{fig:sectors}
\end{figure}

We utilized imaging data from SDSS-DR18, for which we crafted a tailored SQL query to extract metadata aligned with our sector-based methodology. An automated Python script processed this metadata and constructed URLs to download compressed FITS files. After downloading, we centered the celestial objects within 45x45 pixel frames based on their Right Ascension (RA) and Declination (Dec) and then converted the images into PNG format. To prepare the data for CNN analysis, we stacked images from all filters to create a five-channel .npy file and normalized pixel values for uniformity. We applied data augmentation techniques to increase the dataset's diversity and enhance the model's robustness.
\begin{figure}[h]
    \centering
    \includegraphics[scale=0.5]{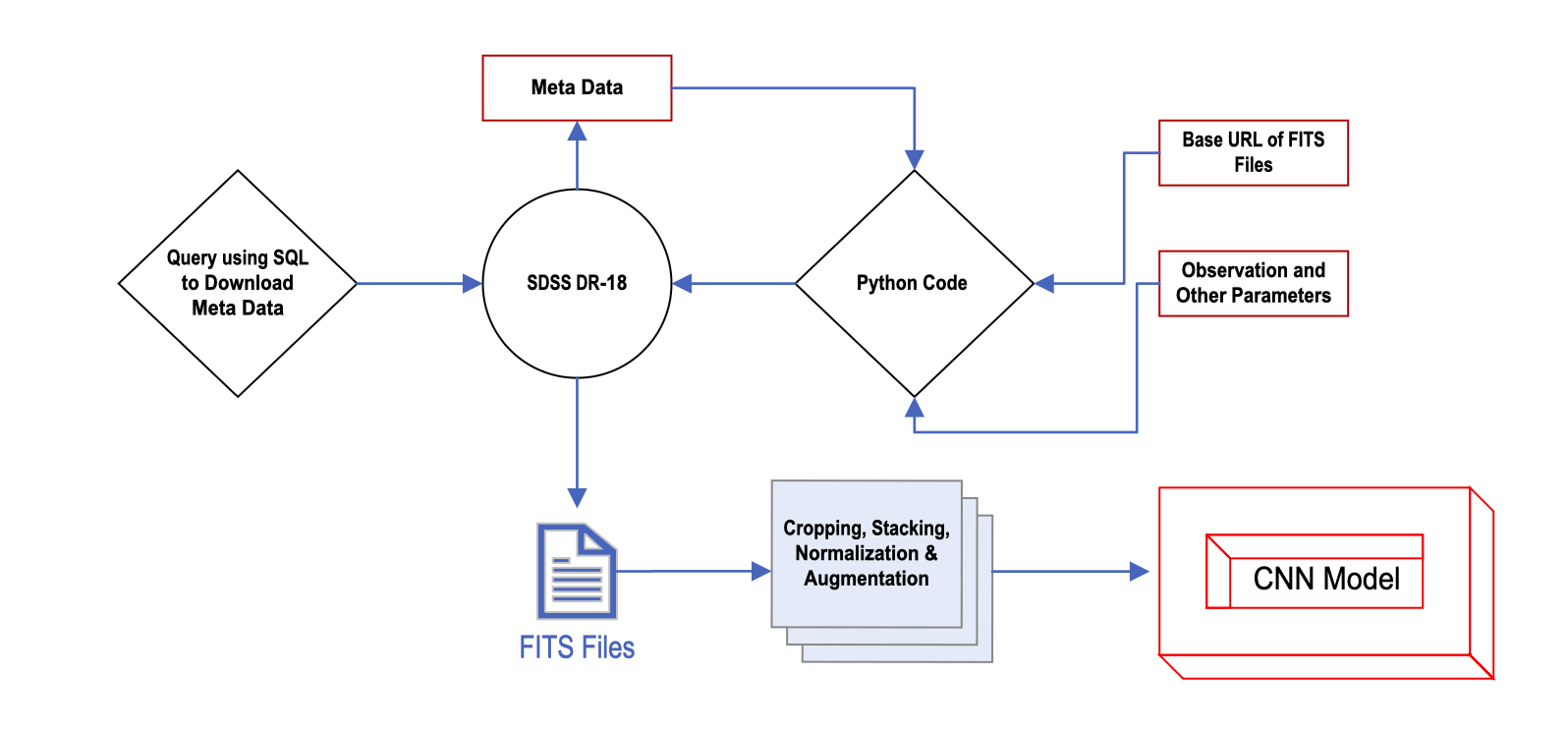}
    \caption{Data Workflow Diagram}
    \label{fig:enter-label}
\end{figure}

\section{Proposed CNN Architecture}\label{appen-2}

\begin{table}[!h]
\centering
\begin{tabular}{|l|l|}
\hline
Layer 1  & Conv2D(32, 3$\times$3 $\times$5),   ReLU , MaxPool \\ \hline
Layer 2  & Conv2D(64, 3$\times$3 $\times$5),   ReLU , MaxPool  \\ \hline
Layer 3  & Conv2D(128, 3$\times$3 $\times$5),   ReLU , MaxPool \\ \hline
Layer 4  & Flatten \\ \hline
Layer 5  & Dense (64) , Dropout(0.5) \\ \hline 
Layer 6  & Dense (1)\\ \hline 
\end{tabular}
\caption{Configuration of the proposed CNN for star-galaxy classification.}
\label{tab:CNN-Arch}
\end{table}

All the networks are trained with a batch size of 32 and an ``Adam'' optimizer with a default learning rate of 0.001. The loss function used for the work is binary cross-entropy keeping in mind the binary nature of the problem.

\begin{table}[h]
\centering
\begin{tabular}{|lccc|}
\hline
\multicolumn{4}{|c|}{\textbf{Computational Cost Comparison}}                                                                                     \\ \hline
\multicolumn{1}{|l|}{Models}            & \multicolumn{1}{l|}{Sector-10}          & \multicolumn{1}{l|}{Sector-16}          & Combined           \\ \hline
\multicolumn{1}{|l|}{\textbf{Proposed}} & \multicolumn{1}{l|}{\textbf{15s/epoch}} & \multicolumn{1}{l|}{\textbf{13s/epoch}} & \textbf{25s/epoch} \\ \hline
\multicolumn{1}{|l|}{CovNet}            & \multicolumn{1}{l|}{80s/epoch}          & \multicolumn{1}{l|}{80s/epo}            & 180s/epoch         \\ \hline
\multicolumn{1}{|l|}{MargNet}           & \multicolumn{1}{l|}{1000s/epoch}        & \multicolumn{1}{l|}{570s/epoch}         & 1610s/epoch        \\ \hline
\end{tabular}
\caption{Comparison of Running Time per Epoch of Proposed and Existing Models}
\label{tab:RunningTime}
\end{table}

\section{Data Splitting for Model Training and Testing}\label{appen-3}
Our dataset consists of 20,000 augmented images, equally distributed across Sector-10 and Sector-16, with each sector containing 10,000 images (5,000 stars and 5,000 galaxies). We applied a train-test split of 0.2.
\begin{itemize}
\item \textbf{Individual Sector Analysis:} In each sector, 8,000 images are used for training, and 2,000 images are used for testing.
\item \textbf{Combined Sector Analysis:} When sectors are combined, the total dataset comprises 20,000 images. Here, 16,000 images are used for training, and 4,000 images are used for testing.
\end{itemize}

\begin{figure}[t]
    \centering
    \includegraphics[scale=0.45]{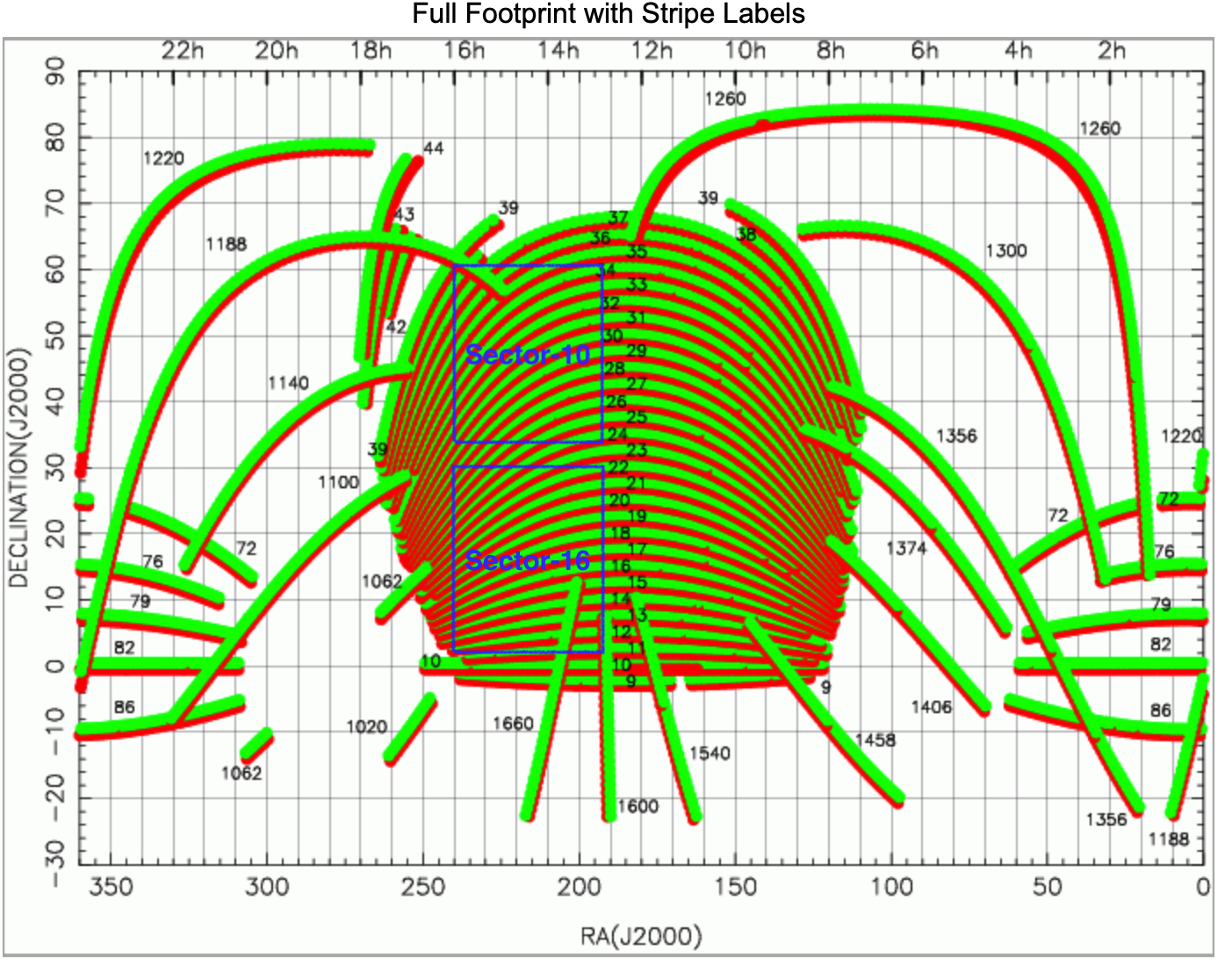}
    \caption{SDSS Sky Coverage \citep{sdss-dr7}}
    \label{fig:ssc}
\end{figure}

\begin{table}[t]
\centering
\caption{Available SDSS Data Across Sectors}
\label{tab:sdssstats}
\begin{tabular}{|c|c|c|c|c|}
\hline
Sector & RA\_Range    & Dec\_Range   & No\_of\_STARS & No\_of\_GALAXIES \\ \hline
1      & 0° to 60°    & 60° to 90°   & 273           & 159              \\ \hline
2      & 60° to 120°  & 60° to 90°   & 5454          & 262              \\ \hline
3      & 120° to 180° & 60° to 90°   & 13133         & 31054            \\ \hline
4      & 180° to 240° & 60° to 90°   & 8239          & 33973            \\ \hline
5      & 240° to 300° & 60° to 90°   & 6628          & 2980             \\ \hline
6      & 300° to 360° & 60° to 90°   & 4771          & 381              \\ \hline
7      & 0° to 60°    & 30° to 60°   & 21537         & 31868            \\ \hline
8      & 60° to 120°  & 30° to 60°   & 26896         & 45414            \\ \hline
9      & 120° to 180° & 30° to 60°   & 125641        & 418122           \\ \hline
10     & 180° to 240° & 30° to 60°   & 113805        & 37085            \\ \hline
11     & 240° to 300° & 30° to 60°   & 49199         & 102023           \\ \hline
12     & 300° to 360° & 30° to 60°   & 20660         & 20119            \\ \hline
13     & 0° to 60°    & 0° to 30°    & 102823        & 306967           \\ \hline
14     & 60° to 120°  & 0° to 30°    & 41578         & 30375            \\ \hline
15     & 120° to 180° & 0° to 30°    & 128893        & 498274           \\ \hline
16     & 180° to 240° & 0° to 30°    & 104402        & 11901            \\ \hline
17     & 240° to 300° & 0° to 30°    & 39779         & 77124            \\ \hline
18     & 300° to 360° & 0° to 30°    & 94264         & 247358           \\ \hline
19     & 0° to 60°    & -30° to 0°   & 58354         & 185367           \\ \hline
20     & 60° to 120°  & -30° to 0°   & 7828          & 1920             \\ \hline
21     & 120° to 180° & -30° to 0°   & 18351         & 28538            \\ \hline
22     & 180° to 240° & -30° to 0°   & 16538         & 41186            \\ \hline
23     & 240° to 300° & -30° to 0°   & 896           & 1144             \\ \hline
24     & 300° to 360° & -30° to 0°   & 31461         & 79353            \\ \hline
25     & 0° to 60°    & -60° to -30° & 0             & 0                \\ \hline
26     & 60° to 120°  & -60° to -30° & 0             & 0                \\ \hline
27     & 120° to 180° & -60° to -30° & 0             & 0                \\ \hline
28     & 180° to 240° & -60° to -30° & 0             & 0                \\ \hline
29     & 240° to 300° & -60° to -30° & 0             & 0                \\ \hline
30     & 300° to 360° & -60° to -30° & 0             & 0                \\ \hline
31     & 0° to 60°    & -90° to -60° & 0             & 0                \\ \hline
32     & 60° to 120°  & -90° to -60° & 0             & 0                \\ \hline
33     & 120° to 180° & -90° to -60° & 0             & 0                \\ \hline
34     & 180° to 240° & -90° to -60° & 0             & 0                \\ \hline
35     & 240° to 300° & -90° to -60° & 0             & 0                \\ \hline
36     & 300° to 360° & -90° to -60° & 0             & 0                \\ \hline
\end{tabular}
\end{table}

\section{Choosing Sectors: An Overview}\label{appen-4}
We strategically chose sectors 10 and 16 for our analysis due to their high star and galaxy counts as seen in below Table \ref{tab:sdssstats}. These sectors are not only dense but also centrally located within the SDSS sky coverage, as seen in Figure \ref{fig:ssc}, making them ideal for developing predictive models. By starting with these sectors, we aim to establish a strong framework that can be implemented in other sectors as well.

\begin{figure}[]
\centering
\includegraphics[width = 0.85\linewidth]{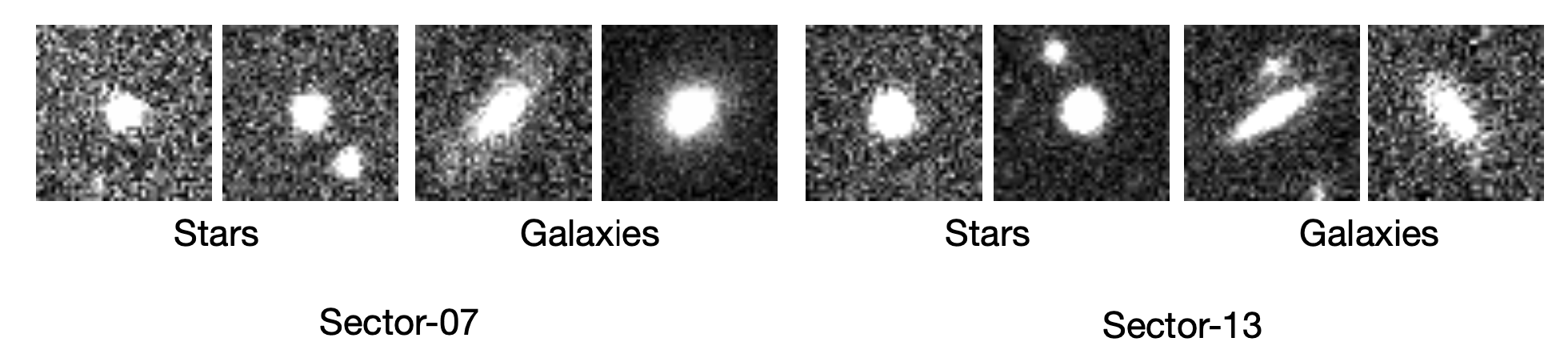}  
\vspace{-3mm}
\caption{A sample image reflecting the challenges in identifying star-galaxies in sectors 7 and 13.}
\label{samples2}
\end{figure}

\begin{figure}[t]
    \centering
\includegraphics[width=0.75\linewidth]{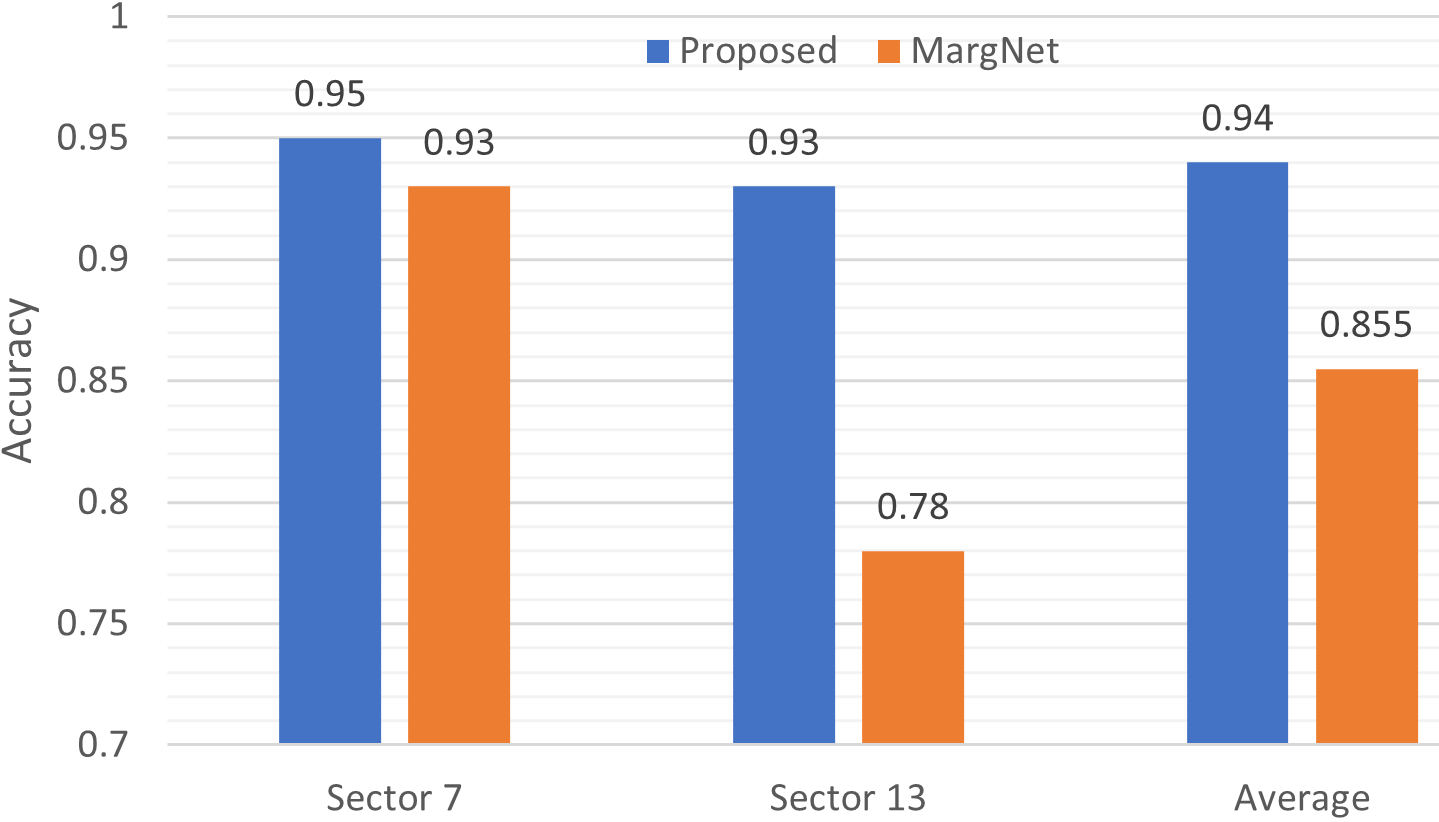}
    \caption{Star-Galaxy Classification performance of the proposed and the MargNet in terms of accuracy (Acc) on Sectors 7 and 13.}
    \label{fig:acc7-13}
\end{figure}

\section{Other Sectors and Zero-Shot Resiliency}

We have also performed experiments in the following manners to evaluate the effectiveness and resiliency of the proposed approach.

\textbf{Evaluation of Sectors 7 and 13:} In this setting, we have downloaded the stars and galaxies images of two new sectors (7 and 13). Similar to setting on sectors 10 and 16, the images of these sectors are divided into training: used to train the model, and testing: used for evaluation. The proposed algorithm yields an accuracy rate of $0.95$ and $0.93$ in comparison to best best-performing existing algorithm namely MargNet, which yields an accuracy of $0.93$ and $0.78$ on sector 7 and 13 images, respectively. The samples of stars and galaxies concerning sectors 7 and 13 are shown in Figure \ref{samples2} and the classification results are reported in Figure \ref{fig:acc7-13}.

\textbf{Zero-shot Sector Resiliency:} In this setting, we have performed a 4-way zero setting experimental evaluation of the proposed algorithm and the MargNet. In other words, we have performed 4 fold cross-validation experiments, where in every fold one unseen sector is used for evaluation, and the remaining three sectors are used for training the classifier.
It is observed that the proposed algorithm yields an average accuracy of $0.89$ in comparison to an accuracy of $0.78$ achieved by the MargNet. Further, the standard deviation of the proposed algorithm and the MargNet are $0.04$ and $0.17$, respectively. We believe the reliability against unseen sectors of the proposed algorithm further establishes the claim of its effectiveness in performing stars and galaxies classification.

\end{document}